# AI-based scoring systematically underestimates conceptual understanding of linguistically weak students' explanations in physics

Markus S. Feser*[1] (ORCID ID: 0000-0001-8503-0951; email: feser@leibniz-ipn.de) &
Paul L. Tschisgale* (ORCID ID: 0000-0001-6763-2457; email: tschisgale@leibniz-ipn.de)

Physics Education, Leibniz Institute for Science and Mathematics Education, Kiel, Germany

*shared lead authorship.

[1]corresponding author

# AI-based scoring systematically underestimates conceptual understanding of linguistically weak students' explanations in physics

## Abstract

Background: Students' explanations of scientific phenomena provide important evidence of their conceptual understanding. However, because conceptual understanding can only be inferred through language rather than observed directly, distinguishing conceptual understanding from linguistic quality represents a fundamental challenge for assessment. This challenge can also be expected to arise when artificial intelligence (AI) is used to score students' text-based explanations.

Results: We investigated whether AI-based scoring approaches assess students' conceptual understanding independently of the linguistic quality of their explanations about a phenomenon taken from a physics context. Conceptual understanding scores assigned by nine machine learning-based scoring approaches and two large language model-based scoring approaches were compared with expert-assigned scores for 116 explanations produced by secondary-school students in Germany. Multinomial logistic regression analyses examined whether expert-rated linguistic quality was associated with under- or overestimation of conceptual understanding. Across all eleven AI-based scoring approaches, lower linguistic quality was associated with a greater likelihood of underestimation, although the strength and statistical significance of this association varied between approaches. A one-unit decrease in linguistic quality was associated with approximately 1.3- to 3.9-times higher odds of underestimation. No comparably consistent association emerged between higher linguistic quality and overestimation.

Conclusions: AI-based scoring approaches may systematically underestimate the conceptual understanding expressed in linguistically weaker explanations. This pattern resembles language bias previously documented in STEM teachers' assessment practices. The findings highlight the importance of examining not only agreement with expert ratings but also whether AI-based scoring approaches inadvertently rely on construct-irrelevant information. Addressing language bias is therefore essential for the validity and fairness of AI-supported assessment in physics education and, potentially, STEM education more broadly.

## 1. Introduction

In STEM education, text-based explanations provide evidence about the scientific quality of students' ideas and their understanding of the causal mechanisms underlying scientific phenomena (Braaten & Windschitl, 2011; Kang et al., 2014). However, such explanations do not offer direct access to understanding; they are linguistic representations through which conceptual ideas are communicated (Schleppegrell, 2004; Wellington & Osborne, 2001). This creates a specific challenge: the linguistic quality of an explanation (i.e. its lexical and syntactic features) can distort inferences about students' conceptual understanding. Assessing text-based explanations therefore requires assessors to carefully distinguish conceptual understanding from linguistic quality. Making this distinction reliably remains a fundamental challenge for assessment in STEM education (Llosa, 2017; Luykx et al., 2007).

This challenge surfaces in STEM teachers' assessment practices. STEM teachers frequently report uncertainty about the role of language in assessing students' text-based explanations and often struggle to determine whether language proficiency should be scaffolded or assessed alongside conceptual understanding (Buxton et al., 2013; Lyon, 2013). Similar challenges have been reported in higher education, where STEM teachers describe difficulties in distinguishing students' language-related difficulties from their understanding of disciplinary content (Wyatt et al., 2021). From an assessment perspective, the entanglement of language proficiency and conceptual understanding is consequential: language proficiency may influence scores assigned by STEM teachers for conceptual understanding, introducing construct-irrelevant variance and potentially disadvantaging students whose scientific understanding is not matched by equally strong linguistic skills (Abedi, 2004; Avenia-Tapper & Llosa, 2015). This concern is consistent with a broader body of research on human assessment showing that the linguistic quality of text-based responses can systematically influence raters' content assessments, even when language is not the intended focus of assessment (e.g. Scannell & Marshall, 1966; Rezaei & Lovorn, 2010; Jansen et al., 2021). In line with that, evidence from STEM education research suggests that teachers' assessments of students' conceptual understanding can be influenced by the linguistic features of their text-based explanations, as demonstrated in physics education (Feser & Höttecke, 2021; Tajmel, 2017) and in research on STEM teachers' interpretations of linguistically diverse student writing (Lemmi et al., 2019; Sedlacek et al., 2025). In the following, we refer to this systematic influence as language bias. As a consequence, language bias raises an important concern about the validity and fairness of STEM teachers' assessments of text-based explanations.

The growing interest in AI-based scoring approaches, which are increasingly being discussed as tools to assist STEM teachers during assessment (e.g. Kortemeyer et al., 2024), amplifies this concern: if such approaches are similarly influenced by linguistic quality, language bias could operate at scale and largely unnoticed. Current AI-based scoring approaches can be broadly divided into machine-learning (ML) approaches and approaches based on large language models (LLMs). ML-based scoring approaches have been researched and used in practice for some time (Zhai et al., 2020a). They typically combine natural language processing, in the form of text-embedding models, with supervised machine learning to automatically assign scores to previously unseen student responses. In contrast, LLMs enable a different approach to AI-based scoring. Rather than learning scoring patterns indirectly from large collections of previously scored responses, LLMs can score students' responses directly from natural-language instructions and examples via prompting (Lee et al., 2024). Despite their different operating principles, it remains unclear—as outlined in the following—whether either approach can infer students' conceptual understanding independently of the linguistic form in which that understanding is expressed.

AI-based scoring approaches are widely advocated in educational assessment on the grounds that they deliver more consistent results than human raters (e.g. Li et al., 2026). Yet consistency offers no protection against systematic error, and a growing body of studies has shown that AI-based scoring exhibits various forms of bias (Baker & Hawn, 2022; Mehrabi et al., 2021). Following Ferrara (2023), bias in the context of AI-based scoring is a systematic error that results in potentially unfair outcomes, where fairness aims to ensure that scoring decisions are equitable and do not disproportionately disadvantage any particular group (Mehrabi et al., 2021). Building on this, an AI-based scoring of a particular construct is considered to be biased if construct-irrelevant features systematically influence the assigned scores. ML-based scoring approaches have repeatedly been shown to encode human-like, stereotyped biases (Caliskan et al., 2017). Likewise, because LLMs are trained on vast amounts of human-generated text, they inevitably absorb societal stereotypes and prejudices alongside useful knowledge (Resnik, 2025), giving rise to a wide range of potential biases (Gallegos et al., 2024). However, with respect to a language bias specifically, the evidence is both more limited and less consistent, and it has largely examined only surface-level features. For ML-based scoring approaches, several studies indicate that surface-level features such as spelling mistakes and minor grammatical errors have little systematic effect on assigned content scores (Ha & Nehm, 2016; Horbach et al., 2017; Kim et al., 2026;). The picture for LLM-based scoring is less settled: Wang et al. (2026) report that LLMs weight linguistic features more heavily than human raters do, whereas Rodrigues et al. (2025) found no difference in content scoring accuracy between texts of native and non-native speakers.

Taken together, research on bias in AI-based scoring approaches has concentrated either on demographic biases or on surface-level linguistic features. The overall linguistic quality of an explanation—its lexical and syntactic sophistication—has scarcely been examined, even though it is precisely this deeper quality that is hardest to separate from the conceptual understanding it expresses. It therefore remains unknown whether the overall linguistic quality of students' text-based explanations systematically influences AI-based scoring of their conceptual understanding—that is, whether the language bias documented for STEM teachers extends to AI-based scoring. To address this research gap, we examined both ML-based and LLM-based scoring approaches, guided by the following research question:

To what extent is the linguistic quality of students' text-based explanations associated with the likelihood of under- and overestimation of their conceptual understanding by AI-based scoring approaches relative to expert-assigned scores?

## 2. Methods

### *2.1 Data*

To address the research question, we used a dataset of authentic student responses for which conceptual understanding and linguistic quality had been scored independently by trained experts. The dataset was originally developed as part of a research project examining how physics teachers assess students' text-based scientific explanations. Details of the original data collection and the development of the associated assessment instruments have been reported elsewhere (Feser, 2019; Feser & Höttecke, 2021).

The final dataset comprised 116 text-based explanations by Grade 9 students from secondary schools in Hamburg, Germany (word count: $M = 35.4$, $SD = 18.5$, median = 33.0). The responses were collected during regular classroom instruction across seven classes from four schools (two academic-track schools and two comprehensive schools). To capture a broad range of students' text-based explanations, schools were purposively selected to represent different educational contexts, including schools with both moderate and comparatively low levels of social disadvantage. All responses were written in German, the language of instruction in the participating physics classrooms.

Students responded individually to the following open-ended physics explanation task (hereafter referred to as the "spacewalk task"; translated from German to English by the authors):

> *"During a spacewalk, the radio connection between two astronauts suddenly fails. Although one astronaut shouts at the top of his lungs, his colleague cannot hear him. The older astronaut grabs his panicking colleague and presses his helmet against the other astronaut's helmet. Suddenly, the younger astronaut can hear the older one faintly. Explain both phenomena in detail."*

The spacewalk task requires students to construct a causal scientific explanation of sound transmission across different physical media and express it in a coherent written account. Rather than recalling isolated facts, students must coordinate conceptual knowledge about sound propagation in vacuum, solids, and gases with causal reasoning and express their understanding in their own written words. Although the dataset was originally developed for research on physics teachers' assessment of students' text-based explanations, it is particularly well suited to the present study. Rather than relying on experimentally manipulated texts, it comprises authentic student explanations that vary naturally in both conceptual understanding and linguistic quality (see Fig. 1). Moreover, because both dimensions were assessed independently using dedicated scoring schemes (described below), the dataset provides an appropriate reference for examining whether AI-based scoring approaches systematically rely on linguistic features when assessing students' conceptual understanding of physics.

### 2.2 Expert ratings

The original dataset included independent expert ratings of both students' conceptual understanding and the linguistic quality of their text-based explanations. These ratings constituted the expert reference used throughout the present study. The complete development and validation of the underlying scoring schemes have been described elsewhere (Feser, 2019). Here, we summarize only those aspects relevant to the present study.

**Development and validation**

The expert reference was developed using a deductive–inductive scoring scheme development process that combined theory-driven category development with iterative analyses of authentic student explanations (Schreier, 2012). The scoring scheme for conceptual understanding was developed by synthesizing existing theoretical frameworks and published scoring schemes for scientific explanations and explanation quality (e.g. Braaten & Windschitl, 2011; Kang et al., 2014), whereas the scoring scheme for linguistic quality synthesized prior work on educational language, scientific language, and language-sensitive STEM education (e.g. Fluck, 1997; Lengyel & Roth, 2012). Both schemes were subsequently refined and calibrated through iterative analyses of authentic student explanations from the spacewalk task until stable and clearly distinguishable categories had been established.

The final expert reference was assigned by one trained physics education research expert, who rated all 116 student explanations using both scoring schemes. To evaluate the reliability and generalizability of these ratings, two independent validation studies were conducted. In the first validation study, a second physics education research expert completed a structured training using authentic student explanations from the spacewalk task to establish a shared interpretation of the scoring schemes before independently rating 31.9% of the dataset. Inter-rater agreement measured by Krippendorff's $\alpha$ was high for both scoring schemes (conceptual understanding: $\alpha = 0.92$; linguistic quality: $\alpha = 0.91$). In the second validation study, six additional physics education research experts completed the same training procedure before independently applying both scoring schemes to a common subset of eight explanations. Agreement across these experts, quantified using Kendall's coefficient of concordance W, demonstrated that the scoring schemes could be applied consistently across independent experts (conceptual understanding: $W = 0.94$; linguistic quality: $W = 0.91$). Together, these development and validation procedures established a reference that operationalized conceptual understanding and linguistic quality as separate constructs while demonstrating high agreement across independent experts.

*Conceptual understanding.* Although the original scoring scheme comprised several complementary content-related dimensions of physics explanations (Feser, 2019), the present study operationalized students' conceptual understanding only using the conceptual consistency dimension of the scheme. This dimension was selected because it directly captures whether the scientific propositions expressed in an explanation form a coherent and scientifically valid account of the underlying physical phenomenon. For the spacewalk task, conceptual understanding was operationalized by assessing whether students' explanations provided a coherent and scientifically valid account of why sound cannot propagate through space but can propagate once the two astronauts press their helmets together, allowing sound to travel through the helmets. Explanations were classified into three ordered categories representing scientifically inconsistent (score = 0), partially consistent (score = 1), or scientifically consistent explanations (score = 2). This classification was explicitly carried out without taking the linguistic quality of the students' text-based explanations into account.

*Linguistic quality*. The scoring scheme for linguistic quality assessed the lexical and syntactic quality of students' explanations written in German as two separate dimensions, irrespective of the scientific correctness of the explanation, with each dimension comprising three ordered categories (scores ranging from 0 to 2). Specifically, the lexical dimension captured the precision, differentiation, and disciplinary appropriateness of students' vocabulary, whereas the syntactic dimension captured how clearly and coherently scientific ideas were communicated through sentence structure, cohesion, and features of scientific written language. Consequently, explanations expressing the same underlying physical idea could receive different linguistic quality scores depending on the way these ideas were linguistically expressed. Spelling and punctuation errors were explicitly excluded because they were not considered indicative of students’ ability to communicate scientific ideas. For the present study, scores for both dimensions of linguistic quality were summed to obtain a single overall measure, resulting in an ordinal score ranging from 0 to 4.

Together, the independently assigned conceptual understanding and linguistic quality scores constituted the expert reference used throughout the present study. Fig. 1 illustrates the distribution and relationship between the expert-assigned scores for conceptual understanding and linguistic quality for the 116 student responses to the spacewalk task. Whereas the conceptual understanding scores served as the reference against which AI-assigned conceptual understanding scores were evaluated, the linguistic quality scores were used exclusively to determine whether deviations between AI-assigned and expert-assigned conceptual understanding scores could be explained by the linguistic quality of students’ text-based explanations.

**Fig. 1.** Distribution and relationship of expert-assigned scores for conceptual understanding and linguistic quality.

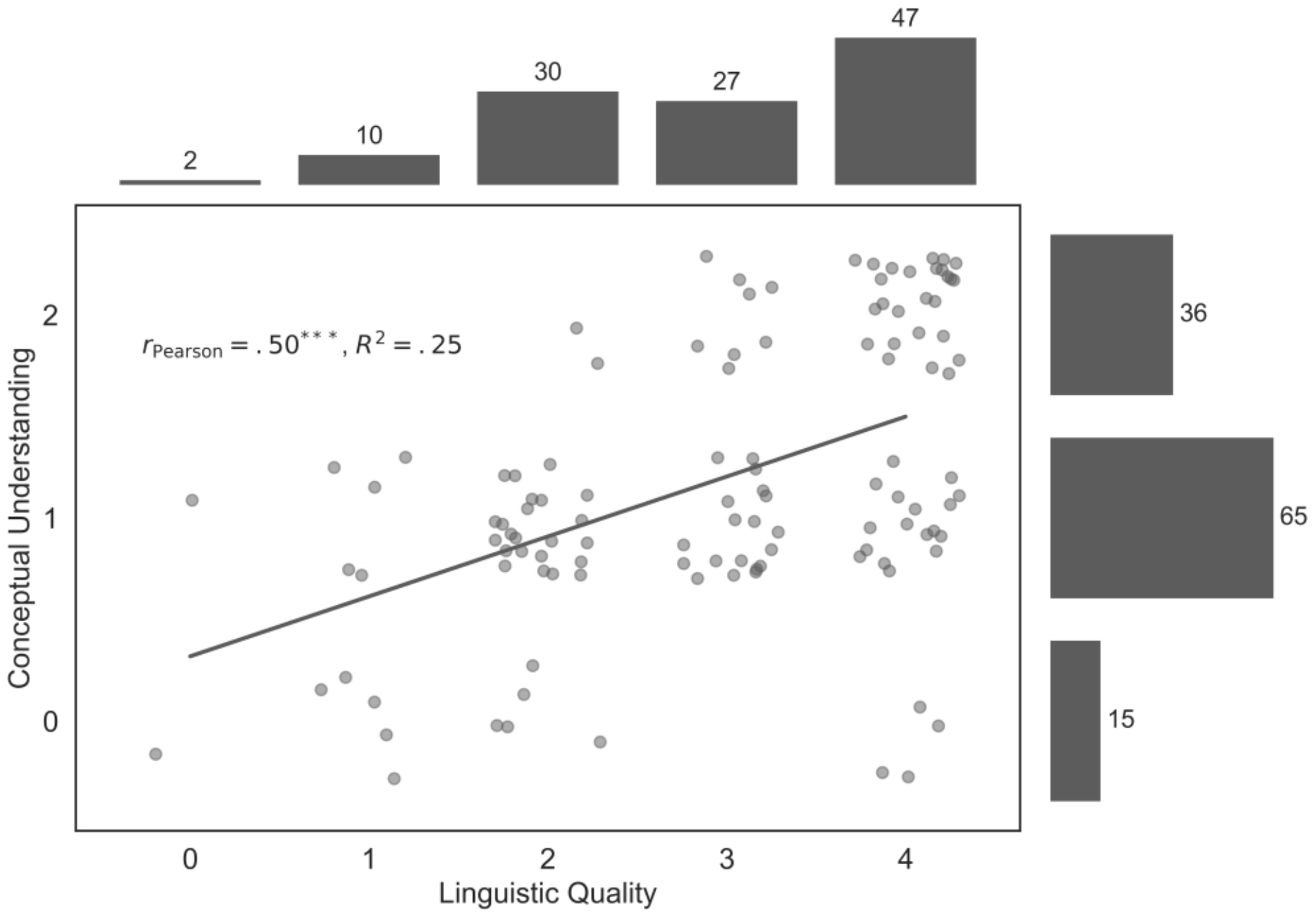

*Note.* Individual observations represent the 116 student responses to the spacewalk task and are displayed as jittered points to reduce overplotting within the discrete score categories. The solid line represents a fitted linear regression and the depicted $R^2$ describes how much variance in conceptual understanding scores is explained by linguistic quality scores. Marginal bar plots show the frequency distributions of the expert-assigned scores for conceptual understanding and linguistic quality.

### *2.3 Using machine learning models to predict conceptual understanding scores*

*Design of the scoring approaches*. We trained multiple ML-pipelines to develop scoring approaches that predict students' conceptual understanding scores based on their responses to the spacewalk task. Each pipeline consisted of two parts: first, a natural language processing step in which a text-embedding model mapped each student text into a high-dimensional vector representing its content; and second, a supervised ML-classifier that mapped this numerical representation onto a conceptual understanding score. We used a fully crossed 3×3 design combining three text-embedding models with three supervised ML-classifiers, yielding nine pipelines that were trained and evaluated separately. The three text-embedding models included a monolingual German model (*German_Semantic_V3*; Huggingface, 2024), a multilingual model (*paraphrase-multilingual-mpnet-base-v2*; Reimers & Gurevych, 2019), and a science-education-specific German model (*G-SciEdBERT*; Latif et al., 2024). The three ML-classifiers were selected to differ in the type of decision boundary they can construct, i.e. the surface in the text-embedding space that separates responses assigned to different conceptual understanding scores. *Quadratic discriminant analysis* (QDA; Tharwat, 2016) produces curved, quadratic boundaries; because it struggles with high-dimensional data, a principal component analysis (PCA) was placed upstream to reduce the embedding dimension. *Regularized multinomial logistic regression* (MLR; Hastie, 2015) constructs simple linear decision boundaries and additionally applies an L1 (lasso) penalty, which effectively selects a small subset of informative embedding dimensions. Finally, the *relevance vector machine* (RVM; Tipping, 1999) is a sparse Bayesian kernel method with non-linear decision boundaries.

*Training and evaluation.* Each pipeline was trained and evaluated within a single 10-fold stratified cross-validation loop (stratified on the three conceptual-understanding scores). Each classifier depends on one or more hyperparameters, i.e. settings that control a model's behaviour and are tuned to improve performance: for QDA, the number of PCA components retained and a hyperparameter stabilizing the per-class covariance estimates; for MLR, the strength of the L1 penalty; and for the RVM, the width of the radial basis function kernel. These were tuned by grid search over predefined ranges of candidate hyperparameter values, each configuration scored by its cross-validated out-of-fold accuracy, and the best-scoring configuration retained per pipeline. For the selected configurations we report accuracy, weighted F1 score, and Cohen's $\kappa$ (with quadratic weights), computed on the same pooled out-of-fold predictions (Additional file 1: Table S1). Hyperparameters were selected on the same cross-validation used to evaluate the final models. This is appropriate here because our goal was not a deployment-ready scoring tool requiring an unbiased estimate of the generalization error, but a well-performing model to serve as the object of the subsequent analysis on language bias. For each of the nine pipelines, the pooled out-of-fold predictions constitute the model-assigned conceptual understanding scores carried into that analysis.

### *2.4 Using LLMs to predict conceptual understanding scores*

*Design of the scoring approach.* As an alternative to the ML-based scoring approaches, we used LLMs to assign conceptual understanding scores directly to the raw text-based explanations of students via prompting, without any specific model training. We tested two OpenAI models, GPT-4.1 and GPT-5-mini (OpenAI, 2024; 2025), each queried through the OpenAI API. Each response to the spacewalk task was appended to a fixed instruction prompt that included the conceptual understanding scoring scheme used for the expert ratings (Feser, 2019; Feser & Höttecke, 2021) and requested the respective LLM to return one of the three conceptual understanding scores as an output. The full prompt written in German (and an English translation) can be found in Additional file 1. Hyperparameters controlling the LLM output were specified a priori: for GPT-4.1 the temperature was set to 1.0; and for the reasoning-optimized model GPT-5-mini, the reasoning effort and text verbosity parameters were both set to "medium." The mapping from text to score is thus carried entirely by the LLMs, as instructed by the prompt and under these fixed settings.

*Evaluation.* Because LLM outputs are stochastic, each student explanation was scored in ten independent runs, i.e. ten separate API calls per response and LLM. This repeated scoring served the purpose of accounting for the run-to-run variability inherent in LLM output generation, rather than resting on a single, potentially idiosyncratic pass. All ten scores per student explanation were treated as separate scores so that reported model performance measured by accuracy, weighted F1 score, and Cohen's $\kappa$ (with quadratic weights) reflects the full spread of outputs across runs (Additional file 1: Table S1).

### 2.5 Evaluating language bias using multinomial logistic regressions

For each student explanation, the conceptual understanding scoring deviation, denoted by $\Delta$, was calculated as the AI-assigned score minus the expert-assigned score. Because conceptual understanding was scored from 0 to 2, deviations could range from $-2$ to 2. Positive deviations indicated overestimation by the AI-based scoring approach, negative deviations indicated underestimation, and a deviation of zero indicated agreement.

For the present study, we focused on the direction rather than the magnitude of the deviation. Scoring outcomes were therefore classified into three categories: underestimation ($\Delta < 0$), agreement ($\Delta = 0$), and overestimation ($\Delta > 0$). Table 1 reports the distribution of these outcomes across all examined AI-based scoring approaches. Notably, the ML-based scoring approaches more often overestimated than underestimated expert-assigned scores, whereas the opposite pattern was observed for the LLM-based systems.

**Table 1.** Proportions of agreement, underestimation, and overestimation across all examined AI-based scoring approaches.

| *ML-based scoring approaches* | | **Agreement** (Δ = 0) | **Underestimation** (Δ < 0) | **Overestimation** (Δ > 0) |
|---|---|---|---|---|
| *Text-embedding model* | *Machine-learning classifier* | | | |
| German_Semantic_V3 | Quadratic Discriminant Analysis | 74.1% | 11.2% | 14.7% |
| | Regularized Multinomial Logistic Regression | 71.6% | 12.9% | 15.5% |
| | Relevance Vector Machine | 75.0% | 9.5% | 15.5% |
| paraphrase-multilingual-mpnet-base-v2 | Quadratic Discriminant Analysis | 77.6% | 6.0% | 16.4% |
| | Regularized Multinomial Logistic Regression | 73.3% | 9.5% | 17.2% |
| | Relevance Vector Machine | 72.4% | 11.2% | 16.4% |
| G-SciEdBERT | Quadratic Discriminant Analysis | 69.8% | 14.7% | 15.5% |
| | Regularized Multinomial Logistic Regression | 78.4% | 9.5% | 12.1% |
| | Relevance Vector Machine | 67.2% | 13.8% | 19.0% |
| ***LLM-based scoring approaches*** | | | | |
| | GPT-4.1 | 72.5% | 23.0% | 4.5% |
| | GPT-5-mini | 74.4% | 15.0% | 10.6% |

*Note.* Δ = AI-assigned score minus expert-assigned score. Agreement indicates identical conceptual understanding scores assigned by the AI-based scoring approach and the human expert (Δ = 0). Underestimation refers to AI-assigned scores lower than the corresponding expert-assigned scores (Δ < 0), whereas overestimation refers to AI scores higher than the corresponding expert-assigned scores (Δ > 0). For GPT-4.1 and GPT-5-mini, proportions are based on all individual scoring runs ($N$ = 1,160 per model).

To evaluate whether the scoring deviations reflected a potential language bias, we examined whether the likelihood of underestimation or overestimation was systematically associated with linguistic quality. A pattern in which explanations with higher linguistic quality were more likely to be overestimated or explanations with lower linguistic quality were more likely to be underestimated would indicate a language bias. Hence, for each AI-based scoring approach, the scoring deviation direction was modelled as an unordered three-category outcome—underestimation (Δ < 0), agreement (Δ = 0), and overestimation (Δ > 0)—using multinomial logistic regression, with agreement specified as the reference category. This parametrization yields two sets of regression coefficients per AI-based scoring approach, contrasting underestimation and overestimation against agreement. Each model regressed the deviation direction on expert-assigned scores for linguistic quality (focal predictor) and expert-assigned scores for conceptual understanding (covariate). Model fit is reported as McFadden's pseudo $R^2$.

Conceptual understanding was included as a covariate because the deviation direction is structurally constrained by it. Since conceptual understanding is scored on a closed 0–2 scale and Δ is defined as the difference of the AI-assigend score and the expert-assigned score, a response rated 2 within the expert rating cannot be overestimated, and a response rated 0 cannot be underestimated. As the expert-assigned score for conceptual understanding increases, the feasible deviations therefore shift from overestimation towards underestimation, which induces a positive association between conceptual understanding and the likelihood of underestimation and a negative association between conceptual understanding and the likelihood of overestimation—irrespective of the linguistic quality of the response. Because expert-assigned scores for conceptual understanding and linguistic quality are moderately correlated in the present data ($r$ = .50, see Fig. 1), an unadjusted model would attribute part of this scale-induced association to linguistic quality: texts of higher linguistic quality also tend to be related to higher conceptual understanding and are therefore a priori more likely to be underestimated and less likely to be overestimated, without any linguistic feature being involved. Linguistic quality would thus be associated with the deviation direction indirectly, as a proxy for conceptual understanding. Including conceptual understanding as a covariate blocks this indirect route, so that the coefficient for linguistic quality reflects only the direct association between linguistic quality and scoring deviation direction.

The described regression-based analysis applies directly to the ML-based scoring approaches, which return a single deterministic conceptual understanding score per student response. The LLM-based scoring approaches, by contrast, are stochastic: repeated scoring of the same response may yield different outcomes. Each student text was therefore scored ten times by both examined LLMs each and the deviation was determined separately for each scoring outcome as the difference between the AI-assigned and the expert-assigned score for conceptual understanding. Because the ten runs are nested within student texts, they do not constitute independent observations, and fitting a single regression model to all 10×116 = 1,160 scoring outcomes would underestimate the standard errors of regression coefficients.

In practice, an LLM assigns a single score to a given text-based explanation of a student, and which of its possible scoring outcomes is realized in that instance cannot be known in advance. The ten runs per text were therefore treated as an empirical distribution of these possible outcomes. Accordingly, 1,000 replicate datasets were generated via a bootstrap-like resampling procedure; within each replicate, one of the ten scoring outcomes was drawn at random and independently for every explanation, so that each replicate dataset comprised exactly one AI-assigned score per explanation and thus satisfied the independence assumption of the regression model. The multinomial logistic regression model specified above was fitted separately to each replicate dataset, and the resulting regression coefficients were combined by Rubin's rules, which partition the uncertainty of a pooled coefficient into a within-replicate and a between-replicate component (Enders, 2010). The pooled coefficients thus estimate the associations that a single scoring run would yield, and their standard errors incorporate both the uncertainty introduced by the finite sample of explanations and the stochasticity of the LLM output. McFadden's pseudo $R^2$ is reported as the mean and standard deviation across the 1,000 replicate regression models.

## 3. Results

Across the examined AI-based scoring approaches, agreement between AI-assigned and expert-assigned conceptual understanding scores ranged from 67.2% to 78.4%, with the remaining disagreements split between underestimations (6.0% to 23.0%) and overestimations (4.5% to 19.0%) (Table 1).

Explanations of lower linguistic quality were systematically more likely to be underestimated (i.e. to receive lower conceptual understanding scores within the AI-based scoring compared to the expert-assigned scores), a pattern observed across all AI-based scoring approaches (Fig. 2). The regression coefficient for linguistic quality was negative for every scoring approach and reached statistical significance ($p < 0.05$) in five of the eleven examined AI-based scoring approaches. Additionally, two scoring approaches reached the more lenient significance threshold ($p < 0.10$; Schumm et al., 2013) commonly adopted in exploratory analyses ($p = 0.053$ and $p = 0.054$; Additional file 1: Tables S2 and S3). Overall, a one-unit decrease in linguistic quality was associated with approximately 1.3- to 3.9-times higher odds of underestimating conceptual understanding ($1.30 \leq 1/OR \leq 3.85$) across all scoring approaches. Despite substantial differences in how these approaches score—ML-based approaches combining several embedding models and classifiers trained on previously scored responses, versus LLM-based approaches scoring directly from natural-language instructions via prompting—the direction of the association remained remarkably consistent, indicating that the observed language bias was not restricted to a specific AI-based scoring approach.

No comparable pronounced association emerged between explanations of higher linguistic quality and overestimation of conceptual understanding across the examined AI-based scoring approaches (Fig. 2). Linguistic quality was unrelated to overestimation in nine of the eleven examined scoring approaches (Additional file 1: Tables S2 and S3). Significant positive associations between linguistic quality and overestimation were observed only for two of the ML-based scoring approaches (both relying on the science education-specific text-embedding model); none of the remaining scoring approaches yielded positive associations that approached either the conventional ($p < 0.05$) or the more lenient ($p < 0.10$) significance threshold.

In summary, the findings indicate that the observed language bias primarily reflected an increased likelihood of underestimating the conceptual understanding represented through text-based explanations of lower linguistic quality rather than overestimating the conceptual understanding represented through text-based explanations of higher linguistic quality. Detailed results of all multinomial logistic regression analyses are provided in Additional file 1: Tables S2 and S3.

**Fig. 2.** Forest plot illustrating the association of linguistic quality with under- and overestimation of conceptual understanding, showing the linguistic-quality coefficients from the multinomial logistic regression analyses.

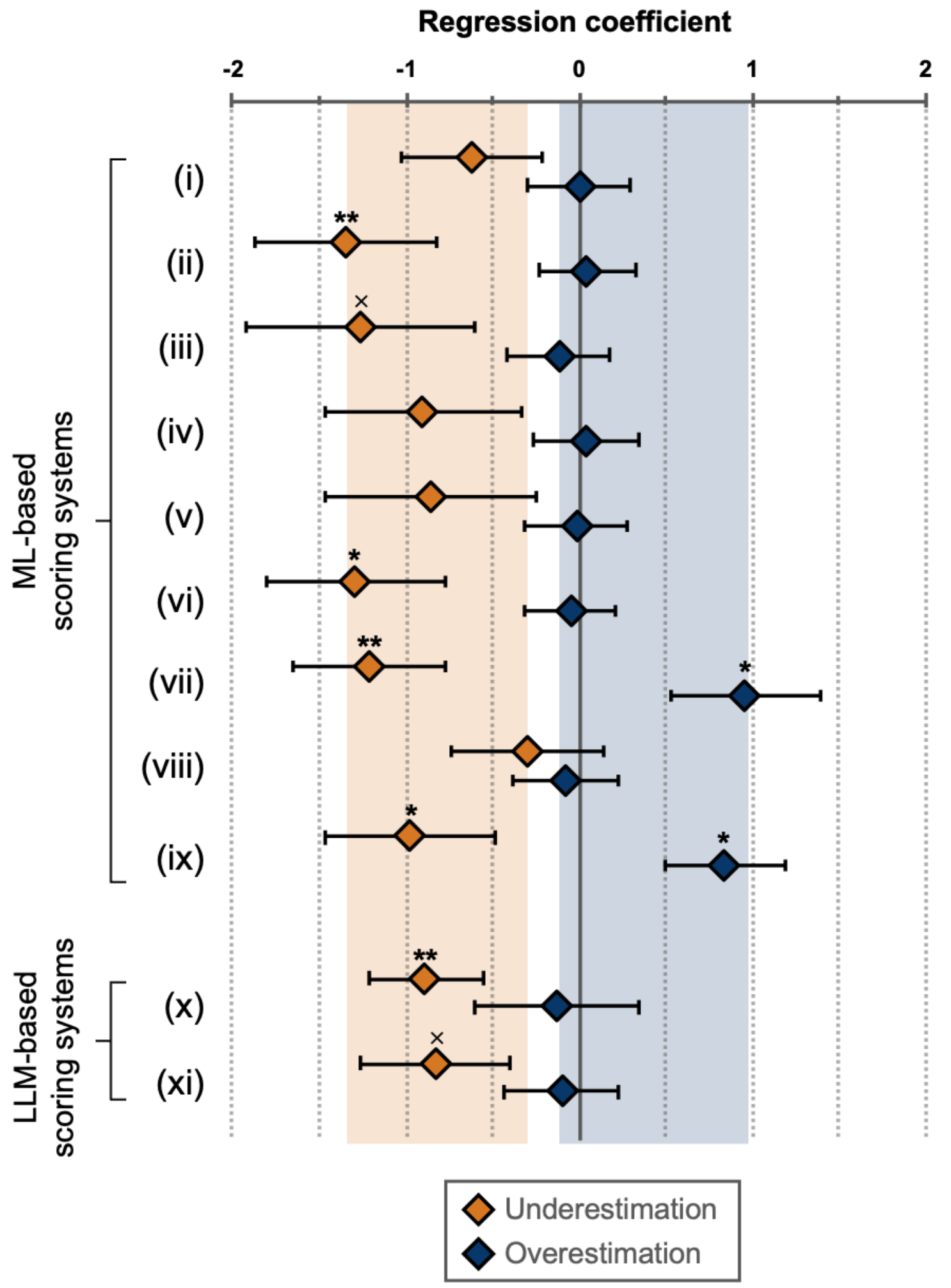

*Note.* The figure summarizes the regression coefficients for linguistic quality; complete model estimates are reported in Additional file 1: Tables S2 and S3. *p*-value labels: ** $p < 0.01$, * $p < 0.05$, × $p < 0.10$. Error bars represent standard errors. Shaded areas indicate the range of regression coefficient estimates for underestimation and overestimation of conceptual understanding across all AI-based scoring approaches. Labels (i)–(ix) denote ML-based scoring approaches combining three text-embedding models (monolingual German model, multilingual model, science-education-specific German model) with three machine learning classifiers (QDA, MLR, and RVM; see Methods). For each embedding model, labels are assigned in classifier order (QDA → MLR → RVM), yielding (i–iii), (iv–vi), and (vii–ix), respectively. Labels (x) and (xi) denote the two LLM-based scoring approaches using GPT-4.1 and GPT-5-mini.

## 4. Discussion

We found consistent evidence that AI-based scoring approaches systematically underestimated students' conceptual understanding when their text-based explanations of a physics phenomenon were expressed in linguistically weaker ways. That this form of language bias emerged across both the investigated ML- and LLM-based scoring approaches suggests that it is not tied to either of the investigated AI-based scoring approaches.

Why language bias primarily takes the specific form of underestimation can be explained by how evidence of conceptual understanding must be inferred from text-based explanations. Such explanations do not provide direct access to students' conceptual understanding; they are linguistic representations through which that understanding is communicated (Schleppegrell, 2004), so that linguistic and conceptual features tend to be closely aligned in authentic student responses (Härtig et al., 2015). Scoring conceptual understanding therefore necessarily relies on identifying evidence of that understanding within the linguistic features of the explanation, and linguistic quality governs how readily such evidence can be identified. When vocabulary and syntax are imprecise, evidence of understanding may be obscured, and conceptual understanding is likely to be underestimated. Conversely, precise use of scientific language makes such evidence easier to identify, making underestimation much less likely. For two of the three ML-based scoring approaches using a science education-specific text-embedding model, however, we also observed that conceptual understanding was overestimated for explanations of higher linguistic quality (Fig. 2, (vii), (ix)). One reasonable mechanism our data cannot test is that the science-specialized embedding model encodes the linguistic features of scientific register in a form that is not clearly separable from the textual cues diagnostic of conceptual quality, so that fluent yet conceptually shallow explanations are represented similarly to genuinely strong ones and the downstream classifier hence scores them too high.

These findings should not be read as an argument against using AI-based scoring approaches in physics education or, more generally, in STEM education. Such approaches can offer considerable opportunities, including scalable scoring of students' text-based explanations, timely formative feedback, and reduced assessment workload for STEM teachers. The stakes of a language bias rise sharply, however, as AI-based scoring approaches move toward summative use—for instance, exam scoring in large undergraduate physics courses (Kortemeyer et al., 2024)—where biased scores bear directly on the validity and fairness of the resulting educational decisions. The concern is most acute for international students and multilingual learners, whose responses may reflect differences in language proficiency rather than conceptual understanding (Rosenthal, 1996), and who would therefore be systematically disadvantaged.

This kind of disadvantage is easily missed, because the standard against which the validity of AI-based scoring approaches is usually evaluated is not designed to detect it. Such approaches are typically evaluated by their agreement with expert ratings (Zhai et al., 2020a), and sufficient agreement is often treated as evidence of valid scoring. Yet agreement is necessary but not sufficient (Williamson et al., 2012): an AI can reproduce expert-assigned scores quite well while its deviations remain systematically tied to construct-irrelevant features (e.g. Zhai et al., 2020b). Our results provide exactly such evidence, with deviations associated with the linguistic quality of students' explanations. Evaluations of AI-based scoring approaches should therefore examine not only the extent to which they agree with expert ratings, but also to what extent disagreements rest on construct-irrelevant information. Providing such evidence is, in turn, essential to establishing valid, equitable, and accountable AI-based assessment for educational decision-making (Grimm et al., 2023).

One natural response to this need is to keep the human in the loop as a safeguard against biased AI-based scoring. Read against the existing literature, however, our findings complicate that response. The striking aspect of our findings is not that AI-based scoring approaches exhibited a language bias, but that this bias closely resembles the language bias previously reported for STEM teachers' assessment of students' text-based explanations. This similarity suggests that the challenge may lie less in the assessor than in the nature of scoring text-based explanations themselves. Because conceptual understanding can only be inferred through the language in which it is expressed, separating conceptual understanding from linguistic quality represents a fundamental challenge for any approach to assess text-based explanations, and any assessor—human or AI-based—may come to rely, at least in part, on linguistic features as indicators of conceptual understanding. Language bias should therefore be understood not as a limitation of particular assessors but also as a more fundamental challenge inherent in the task of inferring conceptual understanding from text-based explanations in STEM education.

Nonetheless, human–AI collaboration may be productive, since the two assessors might not assign biased scores on the same student explanations: a STEM teacher may assess conceptual understanding accurately where an AI-based scoring approach is misled by linguistic quality, or vice versa. Where their scoring diverge, disagreements could serve as diagnostic signals that flag textual explanations for closer inspection, and routing these uncertain cases back to a STEM teacher might de-bias the resulting scores due to increased awareness (Madras et al., 2018). This design would fail, however, precisely where both assessors make the same biased scoring decision. For these cases, a more defensible alternative may be a dedicated AI-based scoring approach that identifies explanations of low linguistic quality and routes them to the STEM teacher with an explicit cue to assess conceptual understanding mindfully independently of linguistic quality. Whether such awareness-raising can overcome the language bias identified in this study remains an open question for future research.

Several limitations should be considered when interpreting our findings. First, our analyses were based on expert ratings as the reference standard. Although these ratings were assigned using independent scoring schemes for conceptual understanding and linguistic quality and extensive rater training, they cannot be regarded as a definitive ground truth. Rather, they represent the best available approximation of students' conceptual understanding while remaining subject to the same fundamental challenge discussed throughout this paper:

conceptual understanding can only be inferred indirectly from students' text-based explanations (Llosa, 2017; Luykx et al., 2007). Additionally, our study focused on text-based explanations in physics produced in German by secondary-school students in response to a single explanation task. Whether similar patterns emerge across different STEM fields, different content areas of physics, age groups, languages, or assessment formats requires further research.

Finally, our findings also point to a broader research agenda for AI-based scoring approaches, particularly in physics education and, more generally, across STEM education. Experimental studies that systematically manipulate the linguistic quality of text-based explanations while holding demonstrated conceptual understanding constant would allow stronger causal inferences about the mechanisms underlying language bias in AI-based scoring approaches. At the same time, explainable AI techniques (e.g. SHAP value analyses; Molnar, 2023) could help identify which linguistic and conceptual features different AI-based scoring approaches rely on when assessing students' text-based explanations in STEM education and why approach-specific patterns of language bias emerge.

Text-based explanations remain one of the most powerful means of inferring students' conceptual understanding in STEM education. Yet our findings show that AI-based scoring approaches exhibit the same language bias—the underestimation of linguistically weaker explanations—previously reported for STEM teachers, suggesting that the difficulty lies less in any particular assessor than in the nature of the scoring task itself. This language bias may be particularly consequential for multilingual and international learners whose lower language proficiency may be misread as weaker conceptual understanding, and they rise as AI-based scoring approaches move toward summative use. Developing AI-based scoring approaches that are robust to linguistic variability while remaining sensitive to the actual evidence of conceptual understanding therefore represents an important challenge for future research.

## 5. Declarations

**Ethics approval and consent to participate:** The present study is based on a secondary analysis of student responses originally collected as part of a research project on physics teachers' assessment practices (Feser, 2019; Feser & Höttecke, 2021). The original data collection was conducted in accordance with the ethical and regulatory requirements applicable at the time in Germany and was approved by the Hamburg school authorities. Participation was voluntary, and student responses were collected using individual participant codes without recording students' names. Written informed consent was obtained from the students' legal guardians, and the participating students also provided their consent to participate.

**Consent for publication:** Not applicable. The manuscript does not contain any identifiable personal information about individual participants.

**Availability of data and materials:** The datasets generated and analysed during the current study are available from the corresponding author on reasonable request, subject to applicable data protection and consent requirements. The complete prompts used for LLM-based scoring, including the original German version and an English translation, are provided in Additional file 1.

**Competing interests:** The authors declare that they have no competing interests.

**Funding:** The authors received no specific funding for this research.

**Authors' contributions:** MSF and PLT contributed equally to the conceptualization, methodology, formal analysis, interpretation of results, and writing of the manuscript. MSF was additionally responsible for the original data collection. Both authors read and approved the final manuscript.

**Acknowledgements:** Not applicable.

## 6. Additional files

### Additional file 1

**File format:** .docx

**Title:** Supplementary Information

**Description:** Supplementary tables reporting the performance of the eleven AI-based scoring approaches and the complete results of the multinomial logistic regression analyses, as well as the full LLM scoring prompt in its original German wording and an English translation.

# Additional file 1

## Supplementary Information

**AI-based scoring systematically underestimates conceptual understanding of linguistically weak students' explanations in physics**

Markus S. Feser & Paul L. Tschisgale

Physics Education, Leibniz Institute for Science and Mathematics Education, Kiel, Germany

**Table S1.** Performance of the AI-based scoring approaches.

| *ML-based scoring approaches* | | Accuracy | F1-Score | Cohen's $\kappa$ |
|---|---|---|---|---|
| *Text-embedding models* | *Supervised ML classifiers* | | | |
| German_Semantic_V3 | Quadratic Discriminant Analysis | 0.741 | 0.732 | 0.614 |
| | Regularized Multinomial Logistic Regression | 0.716 | 0.707 | 0.566 |
| | Relevance Vector Machine | 0.750 | 0.736 | 0.632 |
| paraphrase-multilingual-mpnet-base-v2 | Quadratic Discriminant Analysis | 0.776 | 0.758 | 0.649 |
| | Regularized Multinomial Logistic Regression | 0.733 | 0.715 | 0.534 |
| | Relevance Vector Machine | 0.724 | 0.722 | 0.606 |
| G-SciEdBERT | Quadratic Discriminant Analysis | 0.698 | 0.686 | 0.546 |
| | Regularized Multinomial Logistic Regression | 0.784 | 0.779 | 0.678 |
| | Relevance Vector Machine | 0.672 | 0.658 | 0.526 |
| ***LLM-based scoring approaches*** | | | | |
| | GPT-4.1 | 0.725 | 0.728 | 0.671 |
| | GPT-5-mini | 0.744 | 0.743 | 0.684 |

*Note.* The F1 score is reported as a weighted average (per-category F1 averaged across categories, with each category weighted by how often it occurs in the data). Cohen's $\kappa$ is reported with quadratic weights, penalising disagreements in proportion to the squared distance between categories on the ordinal scale.

**Full LLM prompt used for scoring (original German wording)**

Du sollst Schülertexte von Schüler:innen der 9. Jahrgangsstufe in Deutschland zur folgenden Aufgabe bewerten:

**AUFGABE „Weltraumspaziergang":**

„Bei einem Weltraumspaziergang reißt zwischen zwei Astronauten die Funkverbindung ab. Obwohl der eine Astronaut aus Leibeskräften schreit, hört ihn sein Kamerad nicht. Der ältere Astronaut hält seinen in Panik geratenen Kollegen fest und presst seinen Helm an den des Kollegen. Plötzlich kann der jüngere den älteren leise hören. Erkläre beide Phänomene genau!"

**BEWERTUNGSSYSTEM:**

Erklärungen werden hinsichtlich ihrer internalen (Zusammenhalt betreffend) und externalen Konsistenz (Widerspruchsfreiheit betreffend) eingeschätzt.

- Internale Konsistenz: Ursache-Wirkungsbeziehungen sind lückenlos und erklären die im Fokus der Aufgabenstellung stehenden Phänomene/Sachverhalte.
- Externale Konsistenz: Die Erklärung steht nicht im Widerspruch zu naturwissenschaftlichen Begriffen und Konzepten.

Es gibt drei Kategorien:

**(K1) inkonsistente Erklärung**

(Mindestens 1 Kriterium muss erfüllt sein)

- Bezieht sich auf Phänomene/Sachverhalte, die nicht/kaum im Fokus stehen. Vor allem internal inkonsistent, ggf. auch external inkonsistent.
- Umfasst Begriffe/Konzepte ohne fachlich richtigen Zusammenhang. External inkonsistent.
- Wirkt unzusammenhängend/bruchstückhaft und/oder listet Teilinformationen auf.

**Ankerbeispiele K1:**

- „Im Weltraum werden die Schallwellen anders geleitet. Dadurch dass die Funkverbindung besser wird desto näher man einander ist und im Weltraum ist die Funkverbindung sehr schlecht.“
- „Weil man durch die Schwerkraft im Weltraum sich von Weitem nicht mehr richtig hört aber wenn man dann näher rankommt kann man sich ein bisschen hören, weil man dann auch sehr laut spricht.“
- „Weil diese Helme so nah gegeneinander kommen.“

**(K2) partiell konsistente Erklärung**

(Mindestens 1 Kriterium muss erfüllt sein)

- Erklärung ist teilweise korrekt.

  a) Teile sind external inkonsistent (übergeneralisiert, fachlich falsch).

  und/oder

  b) Teile sind external konsistent, aber internal inkonsistent: fachlich richtige Zusammenhänge, aber irrelevant für die Aufgabenstellung.
- Erklärung ist unvollständig. Teile des Sachverhalts werden nicht aufgeklärt.

**Ankerbeispiele K2:**

- „Im Universum gibt es keinen Sauerstoff, somit auch keine Geräusche. Im Helm schon.“
- „Der Schall breitet sich aus und je näher man aneinander ist desto einen kürzeren Weg hat der Schall, er breitet sich über das Glas des Helmes aus.“
- „Der Schall überträgt sich über die Helme.“

**(K3) konsistente Erklärung**

(Beide Kriterien müssen erfüllt sein)

- Erklärung ist lückenlos und umfassend (internal konsistent) und bezieht sich auf die relevanten Phänomene.
- Erklärung steht nicht im Konflikt mit relevanten naturwissenschaftlichen Begriffen/Konzepten (external konsistent).

**Ankerbeispiel K3:**

- „Im Weltraum gibt es keine Luft die den Schall transportiert. Legt man aber die Helme aneinander, übertragen diese den Schall und werden in Schwingung versetzt.“

**DEINE AUFGABE:**

Wenn ich dir einen Schülertext zur Aufgabe „Weltraumspaziergang“ eingebe, ordne ihn genau einer der drei Kategorien (K1, K2, K3) zu und gib das Ergebnis nur im folgenden JSON-Format aus:

```
{
 "category": "K1" | "K2" | "K3",
 "reasoning": "Kurze Begründung für die Einordnung."
}
```

Berücksichtige dabei nur die inhaltliche Qualität (Konsistenz) der Erklärung. Sprachliche Qualität (Formulierungen, Rechtschreibung etc.) darf keine Rolle spielen.

Beachte außerdem, dass es sich um Erklärungen von Schüler:innen der 9. Jahrgangsstufe in Deutschland handelt und deine Bewertung deren typischen Wissensstand berücksichtigen soll.

Hier ist der entsprechende Schülertext:

**Full LLM prompt used for scoring (English translation)**

You are to assess student texts written by students in the 9th grade in Germany for the following task:

**TASK "Spacewalk":**

"During a spacewalk, the radio connection between two astronauts breaks off. Although one astronaut screams with all his might, his comrade does not hear him. The older astronaut holds onto his panicking colleague and presses his helmet against the colleague's helmet. Suddenly the younger one can hear the older one faintly. Explain both phenomena precisely!"

**SCORING SCHEME:**

Explanations are assessed with regard to their internal consistency (concerning cohesion) and external consistency (concerning freedom from contradiction).

- Internal consistency: Cause-and-effect relationships are seamless and explain the phenomena/facts that are the focus of the task.
- External consistency: The explanation does not contradict scientific terms and concepts.

There are three categories:

**(K1) inconsistent explanation**

(At least 1 criterion must be met)

- Refers to phenomena/facts that are not/barely in focus. Primarily internally inconsistent, possibly also externally inconsistent.
- Comprises terms/concepts without a scientifically correct connection. Externally inconsistent.
- Appears incoherent/fragmentary and/or lists partial information.

**Anchor examples K1:**

- “In space, the sound waves are conducted differently. Because of the fact that the radio connection gets better the closer you are to each other and in space the radio connection is very poor.”
- “Because due to gravity in space you can no longer hear each other properly from far away but when you then come closer you can hear each other a little, because you then also speak very loudly.”
- “Because these helmets come so close together.”

**(K2) partially consistent explanation**

(At least 1 criterion must be met)

- The explanation is partly correct.

  a) Parts are externally inconsistent (overgeneralized, scientifically wrong).

  and/or

  b) Parts are externally consistent but internally inconsistent: scientifically correct relationships, but irrelevant to the task.
- The explanation is incomplete. Parts of the situation are not clarified.

**Anchor examples K2:**

- “In the universe there is no oxygen, hence also no sounds. Inside the helmet there is.”
- “The sound spreads out and the closer you are to each other the shorter the path the sound has, it spreads via the glass of the helmet.”
- “The sound is transmitted via the helmets.”

**(K3) consistent explanation**

(Both criteria must be met)

- The explanation is seamless and comprehensive (internally consistent) and refers to the relevant phenomena.
- The explanation does not conflict with relevant scientific terms/concepts (externally consistent).

**Anchor example K3:**

-  "In space there is no air that transports the sound. But if you put the helmets against each other, these transmit the sound and are set into vibration."

**YOUR TASK:**

When I enter a student text for the task "Spacewalk", assign it to exactly one of the three categories (K1, K2, K3) and output the result only in the following JSON format:

```
{
 "category": "K1" | "K2" | "K3",
 "reasoning": "Short justification for the classification."
}
```

Consider only the substantive quality (consistency) of the explanation. Linguistic quality (phrasing, spelling, etc.) must play no role.

Also note that these are explanations by students in the 9th grade in Germany and your assessment should take their typical level of knowledge into account.

Here is the corresponding student text:

**Table S2.** Full results of the multinomial logistic regression analyses for the effects of conceptual understanding and linguistic quality on under- and overestimation of conceptual understanding across all examined ML-based scoring approaches.

| | | QDA Model | | | | | MLR Model | | | | | RVM Model | | | | |
|---|---|---|---|---|---|---|---|---|---|---|---|---|---|---|---|---|
| | | *β* | *OR* | *1/OR* | *SE(β)* | *p* | *β* | *OR* | *1/OR* | *SE(β)* | *p* | *β* | *OR* | *1/OR* | *SE(β)* | *p* |
| German_Semantic_V3 | *Underestimation* | | | | | | | | | | | | | | | |
| | Intercept | -3.42 | | | 1.14 | 0.003 | -5.32 | | | 1.57 | < 0.001 | -20.70 | | | 67.87 | 0.760 |
| | Conceptual understanding | 2.29 | 9.89 | 0.10 | 0.85 | 0.007 | 4.84 | 126.13 | 0.01 | 1.37 | < 0.001 | 12.23 | $2.04\times10^5$ | 0.00 | 33.97 | 0.719 |
| | Linguistic quality | -0.62 | 0.54 | 1.86 | 0.40 | 0.123 | -1.35 | 0.26 | 3.85 | 0.52 | 0.010 | -1.26 | 0.28 | 3.54 | 0.66 | 0.054 |
| | *Overestimation* | | | | | | | | | | | | | | | |
| | Intercept | 0.75 | | | 0.82 | 0.356 | 0.41 | | | 0.77 | 0.594 | 1.00 | | | 0.81 | 0.217 |
| | Conceptual understanding | -2.87 | 0.06 | 17.63 | 0.68 | < 0.001 | -2.41 | 0.09 | 11.11 | 0.62 | < 0.001 | -2.71 | 0.07 | 14.97 | 0.65 | <0.001 |
| | Linguistic quality | < 0.01 | 1.00 | 1.00 | 0.30 | 0.992 | 0.04 | 1.04 | 0.96 | 0.28 | 0.878 | -0.12 | 0.89 | 1.12 | 0.29 | 0.683 |
| | *McFadden's $R^2$* | | | 0.244 | | | | | 0.295 | | | | | 0.360 | | |
| paraphrase-multilingual-mpnet-base-v2 | *Underestimation* | | | | | | | | | | | | | | | |
| | Intercept | -5.61 | | | 2.08 | 0.007 | -20.83 | | | 72.62 | 0.774 | -4.93 | | | 1.55 | 0.001 |
| | Conceptual understanding | 3.61 | 37.13 | 0.03 | 1.44 | 0.012 | 11.57 | $1.06\times10^5$ | 0.00 | 36.33 | 0.750 | 4.40 | 81.70 | 0.01 | 1.30 | < 0.001 |
| | Linguistic quality | -0.90 | 0.40 | 2.47 | 0.57 | 0.110 | -0.86 | 0.42 | 2.37 | 0.61 | 0.161 | -1.29 | 0.28 | 3.63 | 0.51 | 0.012 |
| | *Overestimation* | | | | | | | | | | | | | | | |
| | Intercept | 1.00 | | | 0.84 | 0.231 | 1.11 | | | 0.82 | 0.178 | 0.19 | | | 0.71 | 0.785 |
| | Conceptual understanding | -3.10 | 0.04 | 22.30 | 0.70 | < 0.001 | -2.95 | 0.05 | 19.09 | 0.68 | < 0.001 | -1.69 | 0.19 | 5.40 | 0.54 | 0.002 |
| | Linguistic quality | 0.04 | 1.04 | 0.96 | 0.30 | 0.899 | -0.02 | 0.98 | 1.02 | 0.29 | 0.958 | -0.05 | 0.95 | 1.05 | 0.26 | 0.847 |
| | *McFadden's $R^2$* | | | 0.309 | | | | | 0.358 | | | | | 0.226 | | |
| G-SciEdBERT | *Underestimation* | | | | | | | | | | | | | | | |
| | Intercept | -3.21 | | | 1.04 | 0.002 | -4.26 | | | 1.35 | 0.002 | -5.44 | | | 1.51 | < 0.001 |
| | Conceptual understanding | 3.48 | 32.57 | 0.03 | 0.99 | < 0.001 | 2.05 | 7.77 | 0.13 | 0.86 | 0.017 | 4.28 | 72.25 | 0.01 | 1.23 | < 0.001 |
| | Linguistic quality | -1.22 | 0.29 | 3.40 | 0.44 | 0.005 | -0.30 | 0.74 | 1.34 | 0.44 | 0.499 | -0.98 | 0.38 | 2.65 | 0.49 | 0.047 |
| | *Overestimation* | | | | | | | | | | | | | | | |
| | Intercept | -0.60 | | | 0.89 | 0.505 | 0.37 | | | 0.80 | 0.640 | -0.38 | | | 0.82 | 0.646 |
| | Conceptual understanding | -4.43 | 0.01 | 83.73 | 1.09 | < 0.001 | -2.46 | 0.09 | 11.76 | 0.65 | < 0.001 | -3.74 | 0.02 | 42.03 | 0.88 | < 0.001 |
| | Linguistic quality | 0.96 | 2.61 | 0.38 | 0.43 | 0.025 | -0.08 | 0.92 | 1.09 | 0.30 | 0.779 | 0.84 | 2.32 | 0.43 | 0.35 | 0.016 |
| | *McFadden's $R^2$* | | | 0.342 | | | | | 0.209 | | | | | 0.342 | | |

*Note.* *β* = regression coefficient, *OR* = exp(*β*) = odds ratio; *SE(β)* = standard error, *p* = *p*-value.

**Table S3.** Full results of the multinomial logistic regression analyses for the effects of conceptual understanding and linguistic quality on under- and overestimation of conceptual understanding across both examined LLM-based scoring approaches.

| | GPT-4.1 | | | | | GPT-5-mini | | | | |
|---|---|---|---|---|---|---|---|---|---|---|
| | *β* | *OR* | *1/OR* | *SE(β)* | *p* | *β* | *OR* | *1/OR* | *SE(β)* | *p* |
| *Underestimation* | | | | | | | | | | |
| Intercept | -1.68 | | | 0.78 | 0.030 | -3.22 | | | 1.14 | 0.005 |
| Conceptual understanding | 2.32 | 10.21 | 0.10 | 0.65 | < 0.001 | 2.79 | 16.31 | 0.06 | 0.94 | 0.003 |
| Linguistic quality | -0.89 | 0.41 | 2.44 | 0.33 | 0.007 | -0.83 | 0.44 | 2.29 | 0.43 | 0.053 |
| *Overestimation* | | | | | | | | | | |
| Intercept | -1.56 | | | 1.24 | 0.209 | -0.58 | | | 0.91 | 0.524 |
| Conceptual understanding | -1.02 | 0.36 | 2.79 | 0.90 | 0.255 | -1.22 | 0.30 | 3.38 | 0.65 | 0.062 |
| Linguistic quality | -0.13 | 0.88 | 1.14 | 0.47 | 0.785 | -0.10 | 0.90 | 1.11 | 0.33 | 0.753 |
| *McFadden's $R^2$: M (SD)* | 0.144 (0.015) | | | | | 0.157 (0.024) | | | | |

*Note. β* = regression coefficient, *OR* = exp(*β*) = odds ratio; *SE(β)* = standard error, *p* = *p*-value. The reported *M* and *SD* of McFadden's $R^2$ summarize the final distribution obtained from *N* = 1,000 bootstrap-like resampling runs, each corresponding to one regression model.